\documentstyle[12pt]{article}
\begin{document}

\title{Wigner's Formulation \\
of Noncommutative Quantum Mechanics
}

\author{
  Akira Kokado\thanks{E-mail: kokado@kobe-kiu.ac.jp},\  
  Takashi Okamura\thanks{E-mail: okamura@ksc.kwansei.ac.jp} \  
  and \ 
  Takesi Saito\thanks{E-mail: tsaito@k7.dion.ne.jp} \\
  \\
 {\small{\it{Kobe International University, Kobe 658-0032, Japan${}^{\ast }$}}} \\
 {\small{\it{Department of Physics, Kwansei Gakuin University, Sanda 669-1337, 
 Japan${}^{\dagger ,\ddagger }$}}}}

\maketitle

\begin{abstract}
When we have noncommutativity among coordinates (or conjugate momenta), we consider Wigner's formulation of quantum mechanics, including a new derivation of path integral formula. We also propose the Moyal star product based on the Dirac bracket in constrained systems.
\end{abstract}

\setlength{\parindent}{1cm}
\newpage

\renewcommand{\thesection}{\Roman{section}.}
\renewcommand{\thesubsection}{\Roman{section}.\arabic{subsection}}
\renewcommand{\theequation}{\arabic{section}.\arabic{equation}}
\setcounter{equation}{0}

\section{Introduction}
\indent

  The idea of noncommutative structure at small length scales is not new. In 1947, Snyder first considered "quantized spacetimes." The idea was that noncommutative spacetimes could introduce an effective cutoff in field theory, in analogy to a lattice. Recently there has been a revival of this idea and many papers have been published [1]. 
Especially interesting is a model of open strings propagating in a constant B field background. Previous studies show that this model is related to noncommutativity of D-branes [2], and in the zero slope limit to noncommutative Yang-Mills theory [3]. An intriguing mixing of UV and IR theories has also been found in the perturbative dynamics of noncommutative field theories [4]. \\
\indent
     In order to describe field theories defined on noncommutative spacetimes, we are better to follow Wigner's formulation [5] in quantum mechanics. Consider a scalar field
 $\phi (x)$, which is expressed in Fourier integral
\begin{equation}
\label{101}
 \phi (x) = \int dp D(p)\exp(ipx).
\end{equation}
We now define a corresponding new function
\begin{equation}
\label{102}
 \hat{\phi }(\hat{x}) = \int dp D(p)\exp(ip\hat{x}),
\end{equation}
where $\hat{x}^\mu $'s obey noncommutative commutation relations
\begin{equation}
\label{103}
 [\hat{x}^\mu , \hat{x}^\nu ] = i\theta ^{\mu \nu }. \qquad (\mu ,\nu =1,\cdots ,n).
\end{equation}
Here  $\theta ^{\mu \nu }$ is a noncommutative real antisymmetric parameter. The Wigner
 representation of $\hat{\phi }(\hat{x})$ is defined by
\begin{equation}
\label{104}
 \phi _{_W}(x) = \int \frac{d^np}{(2\pi )^{n/2}}\textrm{Tr}[\hat{\phi }(\hat{x})\exp(ip(\hat{x}-x))],
\end{equation}
where Tr is defined properly in the text, and we find
\begin{equation}
\label{105}
 \phi _{_W}(x) = \phi (x).
\end{equation}
The correspondence $\phi _{_W}(x)\sim \hat{\phi }(\hat{x})$ is called Weyl-Wigner-Moyal(WWM) correspondence. A product of two fields $(\phi \psi )_{_W}$ is then given by the Moyal star product
\begin{equation}
\label{106}
 (\phi \psi )_{_W} = \phi (x)\star \psi (x) = \phi (x)\exp\left(\frac{i}{2}\theta ^{\mu \nu }
 \stackrel{\gets }{\partial }_\mu \stackrel{\to }{\partial }'_\nu \right)\psi (x')|_{x=x'}.
\end{equation}
Noncommutative field theories are, therefore, constructed in terms of the Moyal star product based on WWM correspondence. \\
\indent
     Contrary to field theories, however, Wigner's formulation of noncommutative quantum mechanics seems to have so far been not considered enough. Wigner's theory is another approach to quantum mechanics, in which the density matrix plays an important role. Here, by noncommutativity we mean noncommutativity among coordinates or conjugate momenta or both. This appears sometimes in constrained systems. One of the most simple examples is the system where a charged particle moves in a so strong constant magnetic field that the kinetic energy is neglected [6]. Such a constrained system is familiar in conventional quantum mechanics, but not in Wigner's formulation [7]. Hence, in the following we would like to consider Wigner's formulation of noncommutative quantum mechanics. First, in Sec.2 we give a new derivation of path integral formula based on Wigner's representation. In Sec.3 several noncommutative quantum theories are considered, including the Moyal star product based on the Dirac bracket.  The final section is devoted to concluding remarks.

\section{Path integral in Wigner's formulation}
\setcounter{equation}{0}
\indent

We consider a transition amplitude 
\begin{equation}
\label{201}
 \langle q_i,t_i | q_{i-1},t_{i-1}\rangle  = \langle q_i| \hat{U}(t_i,t_{i-1}) |q_{i-1}\rangle 
\end{equation}
with
\begin{equation} 
\label{202}
 \hat{U}(t_i,t_{i-1}) = \exp\{-i\hat{H}(t_i-t_{i-1})\},
\end{equation}
where $t_i-t_{i-1}=\epsilon $ is an infinitesimal time-interval and $\hat{H}=\hat{H}(\hat{x},\hat{p})$ is a Hamiltonian operator, which is hermitian but not always written in Weyl-ordered forms.\footnote{For example, the Weyl-ordered form of $p^2q^2$ is given by $(p^2q^2+pq^2p+qp^2q+q^2p^2)/4$. The  form of $(p^2q^2+q^2p^2)/2$ is hermitian but not of the Weyl-ordered form.} Here we have a commutation relation $[\hat{x},\hat{p}]=i$ between one-dimensional coordinate $\hat{x}$ and its canonical momentum $\hat{p}$. In the following, quantities with the hat stand for operators, while those without the hat are c-numbers. \\
\indent
Our new method is based on Wigner's representation of Eq.(\ref{202}), which is defined as  
\begin{equation}
\label{203}
 U_{_W}(t_i,t_{i-1}) = \frac{1}{2\pi }\int d\triangle _i d\xi _i\exp\{-i(\xi _i x_i+\triangle _ip_i)\}
 \textrm{Tr}[\hat{U}(t_i,t_{i-1})\exp i(\xi _i \hat{x}+\triangle _i\hat{p})].
\end{equation}
By using a formula
\begin{equation}
\label{204}
 \exp i(\xi _i \hat{x}+\triangle _i\hat{p})=\exp(i\frac{\triangle _i\hat{p}}{2})\exp (i\xi _i \hat{x})
 \exp(i\frac{\triangle _i\hat{p}}{2}),
\end{equation}
the right-hand side of Eq.(\ref{203}) is reduced to
\begin{eqnarray}
\label{205}
 && {1 \over 2\pi} \int d\triangle _i d\xi _i
   e^{-i(\xi _i x_i+\triangle _i p_i)}
  \int dXdX_1\langle X| \hat{U} |X_1\rangle 
 \langle X_1| e^{i\frac{\triangle _i\hat{p}}{2}}e^{i\xi _i \hat{x}}
 e^{i\frac{\triangle _i\hat{p}}{2}} |X\rangle
\nonumber \\
 &&= \int d\triangle _i e^{-i\triangle _i p_i}
   \langle x_i+\frac{\triangle _i}{2}| \hat{U}
   |x_i-\frac{\triangle _i}{2}\rangle~,
\end{eqnarray}
where we use the relation $\exp(i\Delta_i \hat{p}) |X \rangle
= | X - \Delta_1/2 \rangle$.
If we set
\begin{equation}
\label{206}
 x_i+\frac{\triangle _i}{2}=q_i, \qquad x_i-\frac{\triangle _i}{2}=q_{i-1}
\end{equation}
or
\begin{equation}
\label{207}
 x_i =\frac{q_i+q_{i-1}}{2}, \qquad \triangle _i=q_i - q_{i-1}
\end{equation}
and substitute Eq.(\ref{202}) into the last equation of (\ref{205}), we have
\begin{equation}
\label{208}
 U_{_W}(t_i, t_{i-1}) = \int d\triangle _i \exp (-i\triangle _i p_i)\langle q_i,t_i|q_{i-1},t_{i-1}\rangle .
\end{equation}
\indent
     On the other hand, one may substitute $\hat{U}(t_i,t_{i-1})\cong 1-i\epsilon \hat{H}(\hat{x},\hat{p})$, ($\epsilon =t_i - t_{i-1}$) into Eq.(\ref{203}), i.e.,
\begin{equation}
\label{209}
 U_{_W}(t_i, t_{i-1}) \cong \frac{1}{2\pi }\int d\triangle _id\xi _i \exp\{-i(\xi _ix_i+\triangle _ip_i)\}
 \textrm{Tr}[\{1-i\epsilon \hat{H}(\hat{x},\hat{p})\}\exp i(\xi _i\hat{x}+\triangle _i\hat{p})].
\end{equation}
By using a formula 
\begin{equation}
\label{210}
 \textrm{Tr}[\exp i(\xi _i\hat{x}+\triangle _i\hat{p})]  = 2\pi \delta (\xi _i)\delta (\triangle _i),
\end{equation}
Eq.(2.9) is reduced to
\begin{eqnarray}
\label{211}
 U_{_W}(t_i, t_{i-1}) &\cong &1-i\epsilon H_{_W}(x_i,p_i) \nonumber \\
 &=& \exp(-i\epsilon H_{_W}(x_i,p_i)),
\end{eqnarray}
where $H_{_W}(x_i,p_i)$ is the Wigner representation of $\hat{H}(\hat{x},\hat{p})$ defined by
\begin{eqnarray}
\label{212}
 H_{_W}(x_i,p_i) &=& \frac{1}{2\pi }\int d\triangle _id\xi _i \exp\{-i(\xi _ix_i+\triangle _ip_i)\}
 \textrm{Tr}[\hat{H}\exp i(\xi _i\hat{x}+\triangle _i\hat{p})] \nonumber \\
 &=& H_{_W}(\frac{q_i+q_{i-1}}{2},p_i).
\end{eqnarray}
Here we have substituted the first equation in (\ref{207}), i.e., $x_i$ is the middle point of $q_i$ and $q_{i-1}$.  From the inverse Fourier transform of Eq.(\ref{208}) with (\ref{211}) we get
\begin{equation}
\label{213}
 \langle q_i,t_i|q_{i-1},t_{i-1}\rangle =\frac{1}{2\pi }\int dp_i\exp i\{\triangle _ip_i-\epsilon H_{_W}(\frac{q_i+q_{i+1}}{2},p_i)\}.
\end{equation}
This formula is familiar in the conventional path integral formulation, if $\triangle _i =q_i-q_{i-1}$, 
$\epsilon =t_i-t_{i-1}$ are substituted into Eq.(\ref{213}). \\
\indent
     The transition amplitude between finite time-intervals $t'$ and $t$ is then given by
\begin{eqnarray}
\label{214}
 \langle q',t'|q,t\rangle  &=& \int \langle q',t'|q_n,t_n\rangle dq_n\langle q_n,t_n|q_{n-1},t_{n-1}\rangle dq_{n-1}\cdots dq_1\langle q_1,t_1|q,t\rangle  \nonumber \\
 &=& \int \prod ^n _{i=1}dq_i(t_i)\prod ^{n+1}_{j=1}\frac{dp_i}{2\pi }\exp \sum ^{n+1}_{l=1}i[(q_l - q_{l-1})p_l
 - (t_l-t_{l-1})H_{_W}(\frac{q_l+q_{l-1}}{2},p_l)] \nonumber \\
 &=&\int [\frac{dqdp}{2\pi }]\exp i\int ^{t'} _t d\tau [\dot{q}p-H_{_W}(q,p)].
\end{eqnarray}
This coincides with the conventional path integral formula. However, we should note that the Wigner Hamiltonian $H_{_W}(q,p)$ does not always coincide with the classical Hamiltonian $H_{cl}(q,p)$ which is simply replaced $\hat{x}$, $\hat{p}$ with c-numbers $q$, $p$ in $\hat{H}(\hat{q},\hat{p})$. Of course, $H_{_W}(q,p)$ coincides with $H_{cl}(q,p)$ when $\hat{H}$ is written in Weyl-ordered. \\
\indent
     Our derivation is based on Wigner's representation of the time developing operator (\ref{202}).  It is nothing but the Fourier component of transition amplitude (\ref{201}), as is seen from Eq.(\ref{208}).  Contrary to ours, the conventional derivation is based on the direct calculation of the transition amplitude (\ref{201}) itself.  Since both functions are related with each other through the Fourier integral, both derivations are essentially the same.

\section{Noncommutative quantum mechanics}
\setcounter{equation}{0}
\subsection{Simple examples with constraints}
\indent

     The Wigner formulation is considered in noncommutative quantum mechanics. Let us first consider a simple example: A particle moves on a two-dimensional plane. Its coordinates $\hat{x}_1$, $\hat{x}_2$ and their conjugate momenta $\hat{p}_1$, $\hat{p}_2$  obey noncommutative commutation relations
\begin{eqnarray}
\label{301}
 \left[\hat{x}_i, \hat{x}_j\right] &=& i\epsilon _{ij},\nonumber \\ 
 \left[\hat{x}_i, \hat{p}_j\right] &=& i\delta _{ij},\qquad(i, j=1,2)\nonumber \\
 \left[\hat{p}_i, \hat{p}_j\right] &=& i\epsilon _{ij}.
\end{eqnarray}
An actual example that leads to such relations is given in Ref.[6]. The other cases will be considered later. Here we have neglected a discussion of the intermediary procedure of Dirac quantization for constrained systems, since it is irrelevant to our present purpose. \\
\indent
      Now, whenever we have the commutation relations (\ref{301}), we always obtain constraints
\begin{equation}
\label{302}
 \omega _i = \hat{p}_i - \epsilon _{ij}\hat{x}_j=0, \qquad (i,j=1,2).
\end{equation}
because  $\omega _i$'s are commutable with all of $\hat{x}_i$ and  $\hat{p}_i$, hence such quantities are nothing but c-number constants, which are set to be zeros. Let us then eliminate $\hat{p}_i$ from Eqs.(\ref{301}) and (\ref{302}). We are left with a single commutation relation
\begin{equation}
\label{303}
 [\hat{x}_1, \hat{x}_2]=i. 
\end{equation}
The Hamiltonian operator  $H(\hat{x}_1,\hat{x}_2)$ is now a function of $\hat{x}_1$ and $\hat{x}_2$, and $\hat{x}_2$ plays a role of $\hat{p}_1$. The conventional Wigner formulation is, therefore, valid for noncommutative variables $\hat{x}_1$  and $\hat{x}_2$. The path integral formula (2.14) is also true, if  $q$ and $p$ are replaced with $x_1$ and $x_2$, respectively.  The Wigner function of density matrix $\hat{\rho }$ is defined by
\begin{equation}
\label{304}
 \rho _{_W}(\vec{x}) = \frac{1}{2\pi }\textrm{Tr}[\hat{\rho }\hat{F}(\vec{x})], 
\end{equation}
where $\vec{x}=(x_1, x_2)$ , and
\begin{equation}
\label{305}
  \hat{F}(\vec{x})=\int \frac{d^2\xi }{2\pi }\exp\{i\vec{\xi }(\hat{\vec{x}}-\vec{x})\}.
\end{equation}
The trace of [\ \ \ ] can be calculated, for example,
by using eigenstates $|x_1 \rangle$ of $\hat{x_1}$ , as 
$\int dx'_1\langle x_1'|[\ \ \ ]|x_1'\rangle $. 
From this definition Eq.(\ref{304}) is reduced to
\begin{equation}
\label{306}
 \rho _{_W}(\vec{x})=\int \frac{d\xi _2}{2\pi }\langle x_1+\frac{\xi _2}{2}|\hat{\rho }|x_1-\frac{\xi _2}{2}\rangle \exp(-i\xi _2x_2).
\end{equation}
The Wigner representation of dynamical variable $\hat{A}$ is also defined by 
\begin{equation}
\label{307}
 A_{_W}(\vec{x})=\textrm{Tr}[\hat{A}\hat{F}(\vec{x})] .
\end{equation}
The expectation value of $\hat{A}$ is represented as
\begin{equation}
\label{308}
 \langle \hat{A}\rangle  = \textrm{Tr}[\hat{\rho }\hat{A}]=\int d^2x \rho _W(\vec{x})A_W(\vec{x}).
\end{equation}
A product of dynamical variables $(AB)_{_W}$ is seen to be given by the Moyal star product:
\begin{eqnarray}
\label{309}
 (AB)_{_W} &=& A_{_W}(\vec{x})*B_{_W}(\vec{x}) \nonumber \\
 &=&\exp\left(\frac{i}{2}\epsilon _{ij}\frac{\partial }{\partial \xi _i}\frac{\partial }{\partial \zeta _j}\right)
 A_{_W}(\vec{x}+\vec{\xi })B_{_W}(\vec{x}+\vec{\zeta })|_{\vec{\xi }=\vec{\zeta }=0}.
\end{eqnarray}
\indent
     This is a typical example of Wigner's formulation of noncommutative quantum mechanics. We have used the constraints to eliminate $\hat{p}$'s, and then reduced the remaining coordinates $\hat{x}$'s to the canonical form. Wigner's formulation together with the path integral formula is valid for these coordinates.
     
\subsection{Extension to higher dimensions}
\indent
     Let us extend the above $2$-dimensional space to the  $2N$-dimensional space.  Noncommutative commutation relations are given as
\begin{eqnarray}
\label{310}
 \left[\hat{x}_i, \hat{x}_j\right] &=& i\theta _{ij}, \nonumber \\
 \left[\hat{x}_i, \hat{p}_j\right] &=& ig_{ij}, \qquad (i, j=1,\cdots ,2N) \nonumber \\
 \left[\hat{p}_i, \hat{p}_j\right] &=& i(g\theta ^{-1}g)_{ij},
\end{eqnarray}
where $\theta _{ij}$  ($g_{ij}$)is an antisymmetric (symmetric) regular matrix with real c-numbers. Such an example is given in Ref.[6].   From (\ref{310}) we find constraints
\begin{equation}
\label{311}
 \omega _i = \hat{p}_i + (g\theta ^{-1})_{il} \hat{x}_l = 0, \qquad (i,l=1,\cdots ,2N).
\end{equation}
Eliminating $\hat{p}_i$  $(i=1,\cdots ,2N)$ from Eqs.(\ref{310}) and (\ref{311}), we are left with 
\begin{equation}
\label{312}
 [\hat{x}_i, \hat{x}_j] = i\theta _{ij}, \qquad (i, j=1,\cdots ,2N)
\end{equation}
Let us take the canonical form for $\theta _{ij}$ , i.e.,
\begin{equation}
\label{313}
 \tilde{\theta }=\left(
  \begin{array}{ccccc}
    L_1   &  0    &    0   &  \cdots     &  0  \\
    0     &  L_3  &    0   &  \cdots     &   0  \\
    0     &  0    &   L_5  &            &  0  \\
    \vdots  &  \vdots   &    &  \ddots   &   \vdots   \\
    0     &  0    &      0  &    \cdots  &    L_{2N-1}  \\
  \end{array}
\right)
, \quad
 L_{2n-1}=\left(
  \begin{array}{cc}
    0   &  \tilde{\theta }_{2n-1,2n}  \\
    -\tilde{\theta }_{2n-1,2n}   &  0  \\
  \end{array}
\right)
,
\end{equation}
so that only non-zero components of Eq.(\ref{312}) are expressed as
\begin{equation}
\label{314}
 [\hat{z}_{2n-1}, \hat{z}_{2n}] = i\tilde{\theta }_{2n-1,2n}, \qquad (n=1,\cdots ,N)
\end{equation}
where $\hat{z}'$s are the correspondent canonical variables. Hereafter we normalize $\hat{z}'$s in such a way that 
$\tilde{\theta }_{2n-1,2n}=-\tilde{\theta }_{2n,2n-1}=1$  holds. This means that dynamical variables with odd (even) numbers are commutable with each other, whereas  $\hat{z}_{2n}$ plays a role of the conjugate momentum of $\hat{z}_{2n-1}$. It may be convenient to introduce new notations such as 
\begin{equation}
\label{315}
 \hat{z}_{2n-1}\equiv \hat{Q}_n, \qquad \hat{z}_{2n}\equiv \hat{P}_n, \qquad (n=1,\cdots ,N).
\end{equation}
They obey canonical commutation relations
\begin{equation}
\label{316}
 [\hat{Q}_m, \hat{P}_n]=i\delta _{mn}, \ \ [\hat{Q}_m, \hat{Q}_n]=[\hat{P}_m, \hat{P}_n]=0, \qquad (m,n=1,\cdots N).
\end{equation}

The conventional Wigner formulation is, therefore, valid in the  $2N$-dimensional phase space $(\vec{Q},\vec{P})$.  The path integral formula (\ref{214}) also can be extended to the same phase space by replacing $q$ and $p$ with $\vec{Q}$ and $\vec{P}$, respectively. \\
\indent
     The odd dimensional case can be always reduced to the even dimensional case, because the odd dimensional matrix 
 $\theta _{ij}$ is necessary to have at least one zero eigenvalue.

\subsection{Cases without constraints}
\indent
  Let us consider noncommutative commutation relations of the type
\begin{eqnarray}
\label{317}
 \left[\hat{x}_i, \hat{x}_j\right] &=& i\theta _{ij}, \nonumber \\
 \left[\hat{x}_i, \hat{p}_j\right] &=& ig_{ij}, \qquad (i, j=1,\cdots ,N) \nonumber \\
 \left[\hat{p}_i, \hat{p}_j\right] &=& 0,
\end{eqnarray}
where $\theta _{ij}$  is the same defined in Eq.(\ref{310}), while $g_{ij}$ is a regular constant real matrix which is not always symmetric. Here, there are no constraints among variables. However, Eq.(\ref{317}) can be always rewritten in following forms.  Define $\hat{q}_i$ by
\begin{equation}
\label{318}
 \hat{q}_i = \hat{x}_i + \frac{1}{2}(\theta g^{-1})_{ij}\hat{p}_j,
\end{equation}
so that we have
\begin{eqnarray}
\label{319}
 \left[\hat{q}_i, \hat{q}_j\right] &=& 0, \nonumber \\
 \left[\hat{q}_i, \hat{p}_j\right] &=& ig_{ij}, \qquad (i,j=1,\cdots ,N) \nonumber \\
 \left[\hat{p}_i, \hat{p}_j\right] &=& 0.
\end{eqnarray}
After $g_{ij}$ is transformed into a diagonal form and then the corresponding variables are rescaled, Eq.(\ref{319}) become conventional canonical forms like Eq.(\ref{316}). Hence, in this case we find nothing new.  
      
\subsection{Constraints unsolved}
\indent
     In previous subsections we obtain Wigner's formulation for constrained systems, by solving constraints to eliminate redundant canonical variables. In this subsection we consider the same problem without eliminating redundant variables. \\
\indent
     Let us denote canonical variables on the $2N$ dimensional phase space $\Gamma $ as $(q_i,p_i)$, ($i=1,\cdots ,N$). We are enough to consider only the second class constraints
\begin{equation}
\label{320}
 \omega _\alpha (q,p) \approx 0, \quad \alpha =1,\cdots ,2M(<2N).
\end{equation}
By definition of the second class the Poisson bracket for them is a regular matrix
\begin{equation}
\label{321}
 C_{\alpha ,\beta } \equiv \left\{\omega _\alpha ,\omega _\beta \right\}_P, \quad \det C_{\alpha ,\beta }\not\approx 0.
\end{equation}
If we can solve the constraints, we obtain the ($2N-2M$) dimensional reduced phase space $\Gamma ^*$, which is parametrized by canonical variables $(q_I^{\ *},p_I^{\ *})$, ($I=1,\cdots ,N-M$). The original variables $(q_i,p_i)$, ($i=1,\cdots ,N$) are equivalent to variables $(q_I^{\ *},p_I^{\ *},\omega _\alpha )$, ($I=1,\cdots ,N-M; \alpha =1,\cdots ,2M$). \\
\indent
     We assume the existence of quantum system on the reduced phase space $\Gamma ^{*}$, and denote its Hilbert space as $H^{*}$ and any operator as $\hat{A}^{*}$. The Wigner representation of $\hat{A}^{*}$ on $\Gamma ^{*}$ is given by
\begin{equation}
\label{322}
 A_{_W}^{\ *}(\vec{q}^{\ *},\vec{p}^{\ *}) = \textrm{Tr}^{*}[A^{*}F^{*}(\vec{q}^{\ *},\vec{p}^{\ *})],
\end{equation}
where
\begin{equation}
\label{323}
 \hat{F}^{*}(\vec{q}^{\ *},\vec{p}^{\ *})
= \int \frac{d\vec{\xi }^{\ *}d\vec{\triangle }^{*}}{(2\pi )^{N-M}}
 \exp\{i[ \vec{\xi }^{\ *}\cdot (\vec{\hat{q}}^{\ *}-\vec{q}^{\ *})
     +\vec{\triangle }^{*}\cdot (\vec{\hat{p}}^{\ *}-\vec{p}^{\ *})]\}.
\end{equation}
The expectation value of $\hat{A}^{*}$ is defined as
\begin{equation}
\label{324}
 \langle \hat{A}^{*}\rangle = \textrm{Tr}^{*}[\hat{A}^{*}\hat{\rho }^{*}],
\end{equation}
where $\hat{\rho }^{*}$ is a density operator and its Wigner's function is
\begin{equation}
\label{325}
 \rho _{_W}^{\ *}(\vec{q}^{\ *},\vec{p}^{\ *})=(2\pi )^{M-N}\textrm{Tr}^{*}[\hat{\rho }^{*}\hat{F}^{*}(\vec{q}^{\ *},\vec{p}^{\ *})].
\end{equation}
From Eqs.(\ref{322})-(\ref{325}) we have
\begin{equation}
\label{326}
 \langle \hat{A}^{*}\rangle = \int d\vec{q}^{\ *}d\vec{p}^{\ *}A_{_W}^{\ *}(\vec{q}^{\ *},\vec{p}^{\ *})
 \rho _{_W}^{\ *}(\vec{q}^{\ *},\vec{p}^{\ *})
\end{equation}
\indent
     Our task is to rewrite the expectation value (\ref{326}) in terms of the original phase space variables $(q_i,p_i)$, ($i=1,\cdots ,N$). This is easily obtained by Faddeev-Popov's method as
\begin{eqnarray}
\label{327}
 \langle \hat{A}^{*}\rangle &=&
  \int d\vec{q}^{\ *}d\vec{p}^{\ *}
    d\vec{\omega} \delta^{(2M)} ( \vec{\omega} )
   A_{_W}^{\ *}(\vec{q}^{\ *},\vec{p}^{\ *})
   \rho _{_W}^{\ *}(\vec{q}^{\ *},\vec{p}^{\ *})
\nonumber \\
 &=& \int d\vec{q}^{\ *}d\vec{p}^{\ *}
    d\vec{\omega} \delta^{(2M)} ( \vec{\omega} )
  A_{_W}(\vec{q}^{\ *},\vec{p}^{\ *},\omega _\alpha )
  \rho _{_W}(\vec{q}^{\ *},\vec{p}^{\ *},\omega _\alpha )
\nonumber \\
 &=& \int d\vec{q}d\vec{p}
    d\vec{\omega} \delta^{(2M)} ( \vec{\omega} )
  \left[\det (C_{\alpha \beta })\right]^{1/2}
  A_{_W}(\vec{q},\vec{p})\rho _{_W}(\vec{q},\vec{p}), 
\end{eqnarray}
where $A_{_W}$($\rho _{_W}$) is an arbitrary extension of
$A_{_W}^{\ *}$($\rho _{_W}^{\ *}$) onto $\Gamma $,
which is weakly equivalent to $A_{_W}^{\ *}$($\rho _{_W}^{\ *}$),
so that $A_{_W} \delta^{(2M)} ( \vec{\omega} )
=A_{_W}^{\ *} \delta^{(2M)} ( \vec{\omega} )$.
Though this extension is arbitrary, the expectation value (3.27) is unique. This method is well known to be extensively useful in path integral formula in constrained system, and is trivially working well also in Wigner's formulation. \\
\indent
     Finally let us consider a product of operators $\hat{A}$ and $\hat{B}$. The Wigner representation of $\hat{A}\hat{B}$ is given by the Moyal star product based on the Poisson bracket on $\Gamma ^{*}$:
\begin{eqnarray}
\label{328}
 (AB)_{_W}^{\ *}(\vec{q}^{\ *},\vec{p}^{\ *})&=&A_{_W}(\vec{q}^{\ *},\vec{p}^{\ *})\star B_{_W}(\vec{q}^{\ *},\vec{p}^{\ *}) \nonumber  \\
 &=& A_{_W}(\vec{q}^{\ *},\vec{p}^{\ *})\exp\left[\frac{i}{2}\sum _{I=1}^{N-M}\left(\frac{\stackrel{\gets }{\partial }}{\partial q_I^{*}} \frac{\stackrel{\to }{\partial }}{\partial p_I^{*}}-\frac{\stackrel{\gets }{\partial }}{\partial p_I^{*}}\frac{\stackrel{\to }{\partial }}{\partial q_I^{*}}\right)\right] B_{_W}^{\ *}(\vec{q}^{\ *},\vec{p}^{\ *}).
 \nonumber \\
\end{eqnarray}
Classically, the Poisson bracket on $\Gamma ^{*}$ is extended to the Dirac bracket on $\Gamma $. Hence, it is natural to expect that the Moyal star product based on the Dirac bracket on $\Gamma $ will appear in Eq.(\ref{328}) when it is extended onto $\Gamma $, that is, 
\begin{eqnarray}
\label{329}
 (AB)_{_{DW}}(z) &\equiv& A_{_W}(z)\star _{_D}B_{_W}(z)
\nonumber \\
 &=& A_{_W}(z)\exp\left[\frac{i}{2}
  \left\{z_\mu ,z_\nu \right\}_{_D}
   \frac{\stackrel{\gets }{\partial }}{\partial z_\mu }
   \frac{\stackrel{\to }{\partial }}{\partial z_\nu }\right] B_{_W}(z)
\end{eqnarray}
where we use the condensed notation $(z_\mu )=(q_1,p_1,\cdots ,q_{_N},p_{_N})\in \Gamma $. \\
\indent
    In fact, the Eq.(\ref{328}) becomes
\begin{eqnarray}
\label{330}
 A_{_W}^{*}(\vec{q}^{\ *},\vec{p}^{\ *})\star
   B_{_W}^{*}(\vec{q}^{\ *},\vec{p}^{\ *})
 \approx A_{_W}(z^{*})
   \exp\left[\frac{i}{2}
    \left\{z_\rho ^{\ *},z_\sigma ^{\ *}\right\}_{_D}
    \frac{\stackrel{\gets }{\partial }}{\partial z_\rho ^{\ *}}
    \frac{\stackrel{\to }{\partial }}{\partial z_\sigma ^{\ *}}\right]
         B_{_W}(z^{*}),
 \nonumber \\
\end{eqnarray}
where $z^{*}=(q_I^{\ *},p_I^{\ *},\omega _\alpha \approx 0)$,
($I=1,\cdots ,N-M$; $\alpha =1,\cdots ,2M$), 
since the Poisson bracket on $\Gamma^*$ 
can be replaced by the Dirac bracket
$\left\{z_\rho ^{\ *},z_\sigma ^{\ *}\right\}_{_D}$ on $\Gamma$.
When the constraints are linear functions of $z$, as previous sections,
we have
\begin{eqnarray}
\label{331}
 \textrm{the r.h.s of (\ref{330})} &=& A_{_W}(z^{*})\exp\left[\frac{i}{2}\left\{z_\rho ^{*},z_\sigma ^{*}\right\}_{_D}\frac{\stackrel{\gets }{\partial }}{\partial z_\mu }\frac{\partial z_\mu }{\partial z_\rho ^{*}}\frac{\partial z_\nu }{\partial z_\sigma ^{*}}\frac{\stackrel{\to }{\partial }}{\partial z_\nu }\right]B_{_W}(z^{*}) \nonumber \\
  &\approx& A_{_W}(z)\exp\left[\frac{i}{2}\left\{z_\rho ,z_\sigma \right\}_{_D}\frac{\stackrel{\gets }{\partial }}{\partial z_\mu }\frac{\stackrel{\to }{\partial }}{\partial z_\nu }\right]B_{_W}(z) \nonumber \\
  &=& A_{_W}(z)\star _{_D}B_{_W}(z).
\end{eqnarray}
This shows that the Moyal star product based on the Dirac bracket
on $\Gamma$ is weakly equivalent to the Moyal star product
based on the Poisson bracket on $\Gamma^*$.

\section{Concluding remarks}
\setcounter{equation}{0}
\indent 

Our derivation of path integral formula is based on $U_{W}(t_i,t_{i-1})$, which is Wigner's representation of the time developing operator $\hat{U}(t_i,t_{i-1})=\exp\{-i\hat{H}(t_i-t_{i-1})\}$. Contrary to ours, the conventional derivation is based on the direct calculation of transition amplitude $\langle q_i,t_i|q_{i-1},t_{i-1}\rangle = \langle q_i|\hat{U}((t_i,t_{i-1})|q_{i-1}\rangle $ itself. In this sense our derivation is alternatively new. However, since both functions are related with each other through the Fourier integral, as is seen from Eq.(\ref{208}), both derivations are essentially the same.   \\
\indent 
  We have then considered Wigner's formulation of noncommutative quantum mechanics. Noncommutativity among coordinates $\hat{x}$'s and also momenta $\hat{p}$'s  appears often in constrained systems. First we have used constraints to eliminate $\hat{p}$'s, and then transformed remaining coordinates to canonical forms. Wigner's formulation including the path integral formula has been shown to be applicable for these canonical forms. Secondly we have considered constraints without eliminating the redundant variables. For expectation values of dynamical variables the Fadeev-Popov method has been useful to give unique definitions for them. Finally we have proposed the Moyal star product based on the Dirac bracket for constrained systems.

\setcounter{equation}{0}
\section*{Acknowledgements}

\indent

   We are grateful to Reijiro Kubo for his critical comments. After completion of this work we drew our attention to similar works of Ref.[7].


\end{document}